\documentstyle[12pt,aps]{revtex}%
\newcommand{\bce}{\begin{center}}
\newcommand{\ece}{\end{center}}
\newcommand{\be}{\begin{equation}}
\newcommand{\ee}{\end{equation}}
\newcommand{\bea}{\vspace{0.25cm}\begin{eqnarray}}
\newcommand{\eea}{\end{eqnarray}}

\def\PLA{{Phys. Lett.}  A }
\def\PRL{{Phys. Rev. Lett.} }
\def\PRA{{Phys. Rev.} A }

\def\JMO{Journ. of Mod. Opt.}

\begin{document} \draft

\vskip 3cm

\begin{center}
{\bf {\LARGE Proposal of an experimental scheme for realising a translucent
eavesdropping on a quantum cryptographic channel}}
\end{center}

\vspace{ .25cm}
\begin{center}
{M.Genovese \footnote{ \small  genovese@ien.it}} 
\\[0pt]

Istituto Elettrotecnico Nazionale Galileo Ferraris \\[0pt]
Str. delle Cacce 91 \\[0pt]
I-10135 Torino, Italy
\end{center}

\vspace{ 3.5 cm} {\large  Abstract }
\vskip 0.5 cm
Purpose of this paper is to suggest a scheme, which can be realised with
today's technology and  could be used for entangling a probe to a photon
qubit based on polarisation.  Using this probe a translucent or a coherent
eavesdropping can be performed.
\vskip 1.7cm

In the last years quantum mechanics properties of states are assuming a
technological relevance.
Among different applications of quantum mechanics to technology the
possibility of transmitting absolutely confidential messages has been of
great interest. This is due to the possibility of creating a key for
encoding and decoding  secret messages by transmitting single quanta
between two parties (usually dubbed Alice and Bob).
The underlying principle of quantum key distribution (QKD) is that nature
prohibits gaining information on the state of a quantum system without
disturbing it. Thus possible eavesdropping by a third party (usually dubbed
Eve) can be identified.
Since the original proposal of quantum cryptography \cite{Wiesner}, many
different protocols for this kind of transmission have been suggested
\cite{cryp}.

For example in the BB84 scheme \cite{BB84} single photons are transmitted
from Alice to Bob, preparing them at random in four partly orthogonal
polarisation states (at $0^o$ and $90^o$, $45^o$  and $135^o$ for example).
Bob selects the bases for measuring the photons polarisation at random
too. Then Alice and Bob communicate on a classical channel the bases they
have used (but not the results of course): when they have used the same
base Bob knows what polarisation was selected by Alice and can build a key.
If a spy (Eve) tries to intercept the message, she will inevitably
introduce errors, which Alice and Bob can detect by comparing a subsample
of the generated key.

In Ekert's  protocol \cite{Ekert} entangled pairs are used. Both Alice and
Bob receive one particle of the entangled pair. Then they perform a
measurement choosing  among at least three different  directions.   Again,
Alice and Bob communicate on a classical channel the bases they have used:
if measurements were performed along parallel axes they are used for
generating the secret key. The other measurements can be used for a test of
Bell inequalities. If Eve tries  to eavesdrop, she inevitably affects the
entanglement between the two particles leading to a reduction of the
violation of the Bell inequalities, which allows  Alice and Bob to
recognise the presence of the spy.

Finally, in the B92 protocol \cite{B92}, Alice sends a state chosen between
two non-orthogonal states to Bob, who performs a measurement using a
projection on the subspaces orthogonal to  the two states. Then Bob
publicly informs Alice when he obtained positive results (but not, of
course, which measurement he made). Finally they retain only bits
corresponding to these results. 

Other schemes have also been proposed \cite{oth}. Common characteristics of
all these schemes are the presence of a quantum channel, where different
quantum states can be transmitted, and of a classic channel which is used
for selecting a subsample of the transmitted states and for testing the
presence of an eavesdropper.   If the classical channel is eliminated,  Eve
could simply cut the quantum channel and  substituting herself as Bob for
Alice and as Alice for Bob creating in this way two keys with the two other
parties.

Many different experiments have been realised using the former schemes,
demonstrating the feasibility of QKD up to a distance of many kilometers
\cite{lontano}.      

All of them are based on transmission of single photon states, where the
alphabet is based either on photon polarisation or on photon phase.

Concerning the strategy of Eve for eavesdropping, the simplest one is when
she simply intercepts the state, which Alice has sent. Then she performs  a
measurement on this state and finally she sends a new state to Bob
preparing it according to the result of the measurement.

However, more elaborated strategies have been proposed, where Alice does
not stop the  transmitted state, but causes it to interact with a second
state (an ancilla) and then obtains information about the transmitted qubit
thanks to the result of a measurement on the ancilla. This procedure is
known as translucent eavesdropping \cite{TEV}. 

Finally, Alice can use eavesdropping schemes where her probe interacts with
more than one of Alice's qubits. These schemes are known as coherent or
joint attacks \cite{CA}.

Many interesting studies \cite{ET} have been devoted to understanding
general conditions for obtaining a secure transmission between Alice and
Bob in the presence of an eavesdropper, even using non-ideal channels. 

However, as far as we know, no practically realisable scheme for
translucent or coherent eavesdropping has been proposed yet.

The purpose of this paper is to suggest a scheme which can be realised with
today's technology and  could be used for entangling a probe to a photon
qubit based on polarisation.  Using this probe Eve can then perform a
translucent or a coherent eavesdropping (causing the ancilla  interact with
more than one transmitted qubit).

The scheme of the eavesdropping apparatus, shown in the figure,  is  a
Mach-Zender interferometer with a  Kerr cell on one of the arms inserted on
the quantum channel (see the figure).

The input port of the first beam splitter is fed with a single photon of
vertical polarisation, which splits on the two interferometer arms. On arm
1 it interacts with the transmitted qubit inside the Kerr cell.
  
This has no effect except when both the photons interacting in the cell
have vertical polarisation ($\vert V \rangle \vert V \rangle \rightarrow
\vert V \rangle \vert V \rangle e^{i \phi}$) (in the following $H$ will
denote the horizontal polarisation).

Let us suppose that the transmitted qubit is in the general form:
\be
\vert u \rangle = cos (\theta) \vert H \rangle + sin( \theta ) \vert V \rangle
\ee

If we denote the probe photon with $ \vert p \rangle$, the final state is
in the entangled form:

\be
\vert \Psi \rangle = { 1 \over \sqrt{2}} \left [cos (\theta) \vert H
\rangle \vert p \rangle_1 + sin( \theta ) \vert V \rangle \vert p \rangle_1
e^{(i \phi)}+
i \vert u \rangle \vert p \rangle_2 \right ]
\ee
where the suffixes after the probe $\vert p \rangle$ denotes the path
followed and where we have considered  a 50 \% : 50 \% beam splitter (which
allows Eve to obtain the largest information on Alice-Bob transmission).

We have thus obtained the desired entanglement between the probe and the
transmitted qubit, which can be used for  translucent or coherent
eavesdropping.

The possibility of realising this scheme, and thus the interest of it,
derives by the fact that, although admittedly very difficult,  the Quantum
Non Demolition (QND) detection of a single photon is at present possible
\cite{SI,SM}. QND measurements of { \it welcher Weg} (which path) have
already been achieved 
using 100 meter long optical fiber \cite{QNDexp}. Of course, the
implementation
of the present scheme using such devices would be, even though not impossible
in theory, almost impossible in practice. The
recent discovery of new materials with very high Kerr coupling, could
however allow an easier and more realistic implementation of this scheme.
Two candidates as Kerr cell with ultra-high susceptibility to be used for
this scheme are the Quantum Coherent Atomic Systems (QCAS)  \cite{QCAS,SI}
and the Bose-Einstein condensate of ultracold (at nanoKelvin temperatures)
atomic gas \cite{BEC}.
These are recent great technical improvements which could permit the
realisation of small Kerr cells, capable of large phase shift, even with a
single photon probe. In fact, both exhibit extremely high
Kerr couplings compared to more traditional materials. In particular, the
QCAS is rather a simple system to be realised (for a review see
\cite{Arimondo})  and thus represents an ideal candidate in this role.
Incidentally, one can notice that Kerr coupling can be further enhanced by
enclosing the medium in a cavity \cite{Agarwal}.
The scheme that we  propose in this paper could, in principle, be used with
relatively small phase shifts too. However, the maximal efficiency is
reached when a phase shift of $\pi$ is produced on the probe by a single
photon. Recently a Lukin and Imamoglu's paper has shown that this result
can be effectively reached \cite{LU}. Experiments addressed to single
photon QND, using a Kerr cell, are in progress \cite{G}.

This recent development of high coupling Kerr cell has already been applied
to the proposal of schemes for complete teleportation \cite{TVF},
for generating Schr\"odinger cats and modulating quantum interference
\cite{nos1,nos2}, for generating \cite{nos2} GHZ states \cite{GHZ} and for
realising quantum gates \cite{TVF,Chiara}. 

Incidentally, the use of a Bose condensate could also allow Eve to "stock"
her photon till when the other has reached Bob, using the very low
propagation velocity of light inside a suited Bose condensate (as low as
few tens of meters per second) \cite{BEC}. 

For the sake of exemplification in the following we consider the
application of this scheme to a simple procedure of eavesdropping on a
quantum channel where the BB84 protocol is used. 

Let us begin with the simple example where Bob performs measurement on the
base H,V or in the one at $45^o$ degrees.

If Alice sends $H$ ($\theta = 0$), Bob measures $H$ and Eve has the state 
\be
{\vert p \rangle _1 + i \vert p \rangle _2 \over \sqrt{2} }
\ee
and thus after the second beam splitter the photon will be detected by the
photodector D3. On the other hand, if Alice send $V$, Bob measures $V$ and,
for Eve, only photodetector D4 clicks.
In this case no transmission error is inserted on the quantum line by Eve
presence, while she obtains a perfect identification of the transmitted qubit.
 
On the other hand, if Alice sends a photon $ \vert + \rangle =  {\vert H
\rangle  +  \vert V \rangle  \over \sqrt{2} }$, after the Kerr cell the
entangled state is:
\be
\vert \Psi \rangle = { 1 \over 2 \sqrt{2}} \left [ \vert + \rangle ( [1 +
e^{i \phi}] \vert p \rangle_1 + 2 i \vert p \rangle_2 )+
 \vert - \rangle [1 - e^{i \phi}]  \vert p \rangle_1 \right ]
\ee
where
$ \vert - \rangle =  {\vert H \rangle  -  \vert V \rangle  \over \sqrt{2} }$.

If $\phi=\pi$, Eve has a 50 \% probability of observing the photon at D3 and 
a 50 \% probability of observing the photon at D4.
Also, a 50 \% error on Bob measurement is introduced.

The same situation happens when Alice sends a $\vert - \rangle $ photon.

Altogether, the probability of a successful eavesdropping for Eve is
$p=3/4$, which leads to a capacity of the channel (Alice-Eve) \cite{dover}
\be
I_{AE} = 1 + p log_2 p + (1-p) log_2 (1-p) = 0.189
\ee
As the error rate for Bob is $1/4$, the capacity of the channel Alice-Bob
is $I_{AB} = 0.189$ as well.

Of course, a careful spy will not use such a procedure which produces
asymmetric  errors for the two bases, which can be easily identified.

In general, if Alice sends a generic state $\vert u \rangle$ the
probability that Bob measures $\vert u \rangle$ as well and Eve sees a
photon at D3 is
\be
P^3_{uu} = 1 / 2 [ 1- cos^2 (\theta ) sin^2 (\theta) (1 -cos (\phi)) +
cos^2 (\theta) + cos( \phi) sin^2 (\theta)]
\label{P4uu}
\ee
Similarly, in the same situation Eve could observe a photon at D4 with
probability:
\be
P^4_{uu} = 1 / 2 sin^4 (\theta) (1- cos (\phi))
\ee
With similar notations, if Alice sends the orthogonal state :
\be
\vert v \rangle = cos (\theta) \vert V \rangle - sin( \theta ) \vert H \rangle
\ee
one has
\be
P^3_{vv} = 1 / 2 [ 1- cos^2 (\theta ) sin^2 (\theta) (1 -cos (\phi)) +
sin^2 (\theta) + cos( \phi) cos^2 (\theta)]
\ee
and
\be
P^4_{vv} = 1 / 2 cos^4(\theta) (1 - cos (\phi))
\ee
Finally, the probabilities corresponding to introducing an error on the
Alice-Bob communication are:
\be
P^3_{uv}= P^4_{uv}= P^3_{vu}= P^4_{vu}= sin^2 (\theta) cos^2 (\theta)
sin^2( \phi /2)
\label{P3uv}
\ee

From Eq.s \ref{P4uu}-\ref{P3uv} it follows that if Eve uses the base
bisecting the two used by Alice and Bob and intercepts a fraction $\alpha$
of the transmitted qubits (in order to reduce the errors introduced in Bob
measurements), higher information is obtained. In fact, being the error
rate on Alice-Eve channel
\be
q_{AE}=  ( P_3^{vu} + P_3^{vv} + P_4^{uu} + P_4 ^{uv}) / 2 
\ee
one obtains $ I_{AE}=  0.40 \alpha $ (when $ \phi = \pi $). Furthermore, as
this procedure generates symmetric errors for the two bases, it is much
more difficult for Bob to distinguish if these derive from noise in the
quantum channel or by Eve's presence. 

Let us now analyse the security threshold. The error rate in the Alice-Bob
channel is unsafe if the mutual information in the Alice-Bob channel does
not exceed the minimum of the mutual information in the Alice-Eve and
Eve-Bob channels \cite{Eck2}: 
\be I_{AB} \leq \min ( I_{AE} , I_{EB})
\ee

This means that whenever $I_{AB}$ is greater than either $I_{AE}$ or
$I_{EB}$, then at least in principle there is a way for Alice and Bob to
distribute a string of secret information. On the other hand, if $I_{AB}$
is smaller than $I_{AE}$ or $I_{EB}$, no sifting procedure can make the
transmission safe.

In \cite{Eck2} it was suggested that this condition may be overly cautious,
however in  \cite{Eck3} it has been shown that this is not the case for the
entangled translucent eavesdropping scenario.

In our case the information on the Eve-Bob and Alice-Eve channels are $I_{EB}=
I_{AE}= 0.4 \alpha  $, whilst the error introduced in the Alice-Bob channel
becomes $q_{AB} = \alpha ( P_3^{vu} + P_3^{uv} + P_4^{uv} + P_4 ^{vu}) / 2
= \alpha / 4 $.

Thus, the transmission cannot be considered safe if Eve intercepts a
fraction $\alpha=0.755$, or larger, of the transmitted photons.
This corresponds to an error rate on the Alice-Bob channel $ q_{AB} = 0.189$.
Of course, this error is relatively large, but one should not forget that
nowadays quantum channels, both in fibers or air, have huge losses. 
Furthermore, with other protocols this value could be smaller.

If Ekert or BB92 protocols are used, a similar analysis can be carried out.
Many papers with general results about eavesdropping and security of
quantum channels are already published \cite{ET}. We refer to them for
general discussion about application of translucent or joint eavesdropping
and  conditions for obtaining a safe communication in the presence of
eavesdropping.

In conclusion, we think that the proposed scheme represents an interesting
chance for an experimental realisation of eavesdropping on a quantum
channel and, as far as we know, this is the first proposal for a practical
realisation of translucent eavesdropping. Of course, this scheme could also
be used in any other situation where interaction with a qubit (represented
by polarisation properties of a photon) and an ancilla qubit is required.
 
\vskip 1cm
{\bf Acknowledgements}

\noindent We would like to acknowledge the support of MURST via special
programs "giovani ricercatori" Dip. Fisica Teorica Univ. Torino and of
Istituto Nazionale di Fisica Nucleare. Thanks are due to C. Novero for
useful discussions.

{  
}
\vfill \eject

\newpage
{\bf Figure Caption}

The Alice - Bob cryptographic channel, with the eavesdropping apparatus of
Eve, which is constituted of a Mach-Zender interferometer and  a Kerr cell.
 A probe photon can follow the arm of the interferometer where the Kerr
cell is posed or the other. If the transmitted photon has a vertical
polarisation the probe photon phase is changed. The observation of the
probe photon at photodetector D3 or D4 gives Eve information on the
transmitted photon polarisation.


\noindent {\bf References}
\begin{enumerate}

\bibitem{Wiesner} S. Wiesner, Sigact News 15 (1983) 78.

\bibitem{cryp} see for example S.J. Lomonaco, quant-ph 9811056 and ref.s
therein. See also "Special issue of Quantum Communication", Journ. of Mod.
Opt. 41, 12 (1994).

\bibitem{BB84} C.H. Bennet and G. Brassard, Proc. of  Int. Conf. Computer
Systems an Signal Processing, Bangalore (1984) 175.  

\bibitem{Ekert} A.K. Ekert, \PRL 67 (1991) 661.

\bibitem{B92} C. H. Bennet, \PRL 68 (1992) 3121.

\bibitem{oth} see for example: M. Czachor, quant-ph 9812030; A. K. Ekert et
al., \PRL 69 (1992) 1293; K. Shimizu and N. Imoto, \PRA 60 (1999) 157; W.
Tittel, H. Zbinden and N. Gisin, quant-ph 9912035.
 
\bibitem{lontano} W.T. Buttler et al., quant-ph 0001088; W. Tittel et al.,
quant-ph 9911109; H. Zbinden Appl. Phys. B 67 (1998) 743, W.T. Buttler et
al, Phys.Rev.Lett. 81 (1998) 3283; A.V. Sergienko et al., \PRA 60 (1999)
R2622 and ref.s therein.

\bibitem{TEV} see C.a. Fuchs and A. Peres, \PRA 53 (1996) 2038;  A.K.Ekert
et al., \PRA 50 (1994) 1047; H.E. Brandt et al., \PRA 56 (1997) 4456 and
ref.s therein. 

\bibitem{CA} see C. Bennet et al., quant-ph 9611006; J.I. Cirac and N.
Gisin \PLA 229 (1997) 1 and ref.s therein.

\bibitem{ET} S.M. Barnett et al., \JMO 40 (1993) 2501; S.M. Barnett and
S.J.D. Phoenix, \PRA 48 (1993) R5, \JMO 40 (1993) 1443; D. Deutsch et al.,
\PRL 77 (1996) 2818; B. Huttner and A.K. Ekert, \JMO 41 (1994) 2455; 
K.J. Blow and S.J.D. Phoenix, \JMO 40 81993) 33; G. Brassard et al,
quant-ph 9906074, quant-ph9911054; E. Biham and T. Mor, \PRL 78 (1997)
2256, \PRL 79 (1997) 4034. 

\bibitem{QNDexp} M.D. Levenson et al., \PRL 57 (1986) 2473; N. Imoto et al.,
Opt. Comm. 61 (1987) 159. 

\bibitem{SI} H. Schmidt and A. Imamoglu, Opt. Lett. 21 (1996) 1936.

\bibitem{SM} B.C. Sanders and G.J. Milburn, \PRA 39 (1989) 694.

\bibitem{QCAS} U. Rathe et al., \PRA 47 (1993) 4994.

\bibitem{BEC} L. Vestergaard Hau et al., Nature 397 (1999) 594.

\bibitem{Arimondo} E. Arimondo, in Progress in optics XXXV, E. Wolf
editor, Elsevier Science 1996, pag. 257.


\bibitem{Agarwal} G.S. Agarwal, Opt. Comm. 72 (1989) 253.

\bibitem{LU} M.D. Lukin and Imamoglu, \PRL 84 (2000) 1419.

\bibitem{G} G. Brida et al, private communication.

\bibitem{TVF} D. Vitali et al., quant-ph 0003082.

\bibitem{nos1} M. Genovese and C. Novero, Phys. Rev. A 61 032102 (2000).
 
\bibitem{nos2} M. Genovese and C. Novero,  \PLA to appear.

\bibitem{GHZ} D.M. Greenberger et al., \PRA 82 (1999) 1345 and ref.s therein.

\bibitem{Chiara} G. M. D'Ariano, C. Macchiavello, and L. Maccone,
Fortschr. Phys. {\bf 48}, 573
(2000).
 
\bibitem{dover} R.B. Ash, Information Theory, (Dover, New York, USA 1990).

\bibitem{Eck2} A. K. Eckert et al, \PRA 50 (1994) 1047.

\bibitem{Eck3} H.E. Brandt et al, \PRA 56 (1997) 4456.


\end{enumerate}
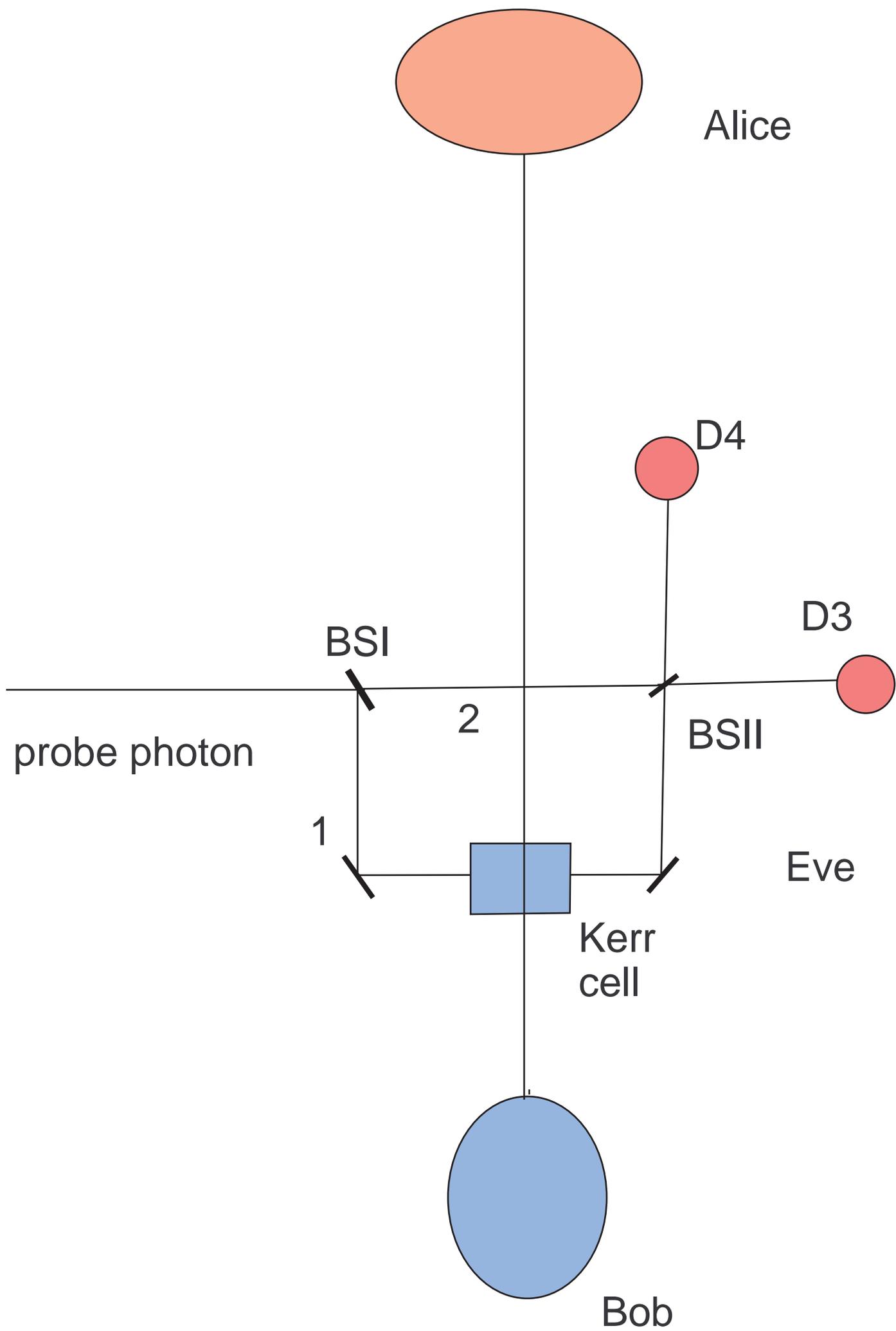
\end{document}